%Paper: gr-qc/9508032
%From: ESPOSITO@napoli.infn.it
%Date: Mon, 14 AUG 95 17:33:37 +0000

\magnification \magstep1
\raggedbottom
\openup 4\jot
\voffset6truemm
\headline={\ifnum\pageno=1\hfill\else
\hfill {\it The Effect of Boundaries in One-Loop Quantum Cosmology}
\hfill \fi}
\rightline {DAMTP R-90/20}
\vskip 1cm
\centerline {\bf THE EFFECT OF BOUNDARIES IN ONE-LOOP}
\centerline {\bf QUANTUM COSMOLOGY$^{*}$}
\vskip 1cm
\centerline {\bf Peter D. D'Eath$^{(a)}$ and Giampiero
Esposito$^{(a,b)}$}
\centerline {$^{(a)}$Department of Applied Mathematics and Theoretical
Physics}
\centerline {Silver Street, Cambridge CB3 9EW, U. K.}
\centerline {$^{(b)}$St. John's College, Cambridge CB2 1TP, U. K.}
\vskip 1cm
\centerline {September 1990}
\vskip 1cm
\noindent
{\bf Abstract.}
The problem of boundary conditions in a supersymmetric theory of quantum
cosmology is studied, with application to the one-loop prefactor in the
quantum amplitude.
Our background cosmological model is flat Euclidean
space bounded by a three-sphere, and our calculations are based on the
generalized Riemann $\zeta$-function. One possible set of supersymmetric
local boundary conditions
involves field strengths for spins 1, ${3\over 2}$
and 2, the undifferentiated spin-${1\over 2}$ field, and a mixture of
Dirichlet and Neumann conditions for spin 0. In this case the results we
can obtain are : $\zeta(0)={7\over 45}$ for a complex scalar field,
$\zeta(0)={11\over 360}$ for spin ${1\over 2}$, $\zeta(0)=
-{77\over 180}$ (magnetic) and ${13\over 180}$ (electric) for spin $1$, and
$\zeta(0)={112\over 45}$ for pure gravity when the linearized magnetic
curvature is vanishing on $S^3$. The $\zeta(0)$ values for gauge fields
have been obtained by working only with physical degrees of freedom.
An alternative set of
boundary conditions can be motivated by studying transformation properties
under local supersymmetry; these involve Dirichlet conditions for the
spin-2 and spin-1 fields, a mixture of Dirichlet and Neumann conditions
for spin-0, and local boundary conditions for the spin-${1\over 2}$ field
and the spin-${3\over 2}$ potential. For the latter one finds :
$\zeta(0)=-{289\over 360}$. The full $\zeta(0)$ does not vanish in
extended supergravity theories, indicating that supersymmetry is one-loop
divergent in the presence of boundaries.
\vskip 12cm
\noindent
$^{*}$Published in Proceedings of
the IX Italian National Congress of General Relativity
and Gravitational Physics, Capri, 25-28 September 1990
(World Scientific Publishing Co.).
\vskip 20cm
This research is concerned with the problem of the one-loop
finiteness of amplitudes in quantum cosmology in the presence of
boundaries. The calculations are performed using the generalized
Riemann $\zeta$-function, working in a particular background with specified
boundary conditions. The value $\zeta(0)$ at the origin provides
information about one-loop divergences of the quantum
amplitudes with the prescribed boundary conditions. One then has to
check whether the contributions to $\zeta(0)$ from bosonic and fermionic
fields add up to zero for a suitable supergravity model.

We consider a background
cosmological model given by flat Euclidean space bounded by a three-sphere
(hereafter referred to as $S^3$). For fermionic fields one has a choice of
local and non-local boundary conditions because of the first-order nature
of the Dirac operator. The former are of greater interest because motivated
by supersymmetry. Using two-component spinor notation, one possible set of
local boundary conditions, involving field strengths for a spin-$s$ field
and the normal to $S^3$ is [1] :
$$
2^{s} \; n^{AA'}...n^{LL'}\phi_{A...L}=\pm {\widetilde \phi}^{A'...L'}
\; \; \; \; .
\eqno (1)
$$
In (1), $n^{AA'}=n^{a}\sigma_{a}^{AA'}$ is the spinor version of the normal
to $S^3$, $\phi_{A...L}$ and ${\widetilde \phi}_{A'...L'}$ are field
strengths not related by complex conjugation. For a complex scalar field,
the real part obeys a Dirichlet (D) and the imaginary part a Neumann (N)
condition (or viceversa). Supersymmetry plays a role
in (1) because in the case of a flat Euclidean background bounded by $S^3$
there is a spin-lowering operator for solutions to the linearized massless
free-field equations which preserves these local boundary conditions [2].
This generates rigid supersymmetry
transformations among classical solutions obeying (1) on
$S^3$. However, these rigid transformations do not map {\it eigenfunctions}
of the spin-$s$ wave operators to eigenfunctions for adjacent spin
$s \pm {1\over 2}$ with the {\it same} eigenvalues. Thus there is no
{\it a priori}
reason for $\zeta(0)$ values for adjacent spins to be equal; the $\zeta(0)$
value for each spin must be calculated separately.
For bosonic fields, conditions (1)
imply the vanishing on $S^3$ of the magnetic (B) or electric (E) field
for spin $1$, whereas for pure gravity (PG) the linearized
magnetic curvature is vanishing on $S^3$ (the fixing of the linearized
electric curvature would lead to an ill-posed classical
boundary-value problem).
For all gauge fields we shall be here concerned
with the method of reduction of the theory to its physical degrees of
freedom (hereafter referred to as PDF), so as to complete previous work
appearing in the literature for bosonic fields [3]. Results obtained
using a quantization technique based on the Faddeev-Popov method can be
found in [4,5]. The PDF results for bosonic fields (compare with [4,5]) are :
$\zeta_{B}(0)=-{77\over 180}, \;
\zeta_{E}(0)={13\over 180}, \;
\zeta_{PG}(0)={112\over 45}$,
and for a complex scalar field one finds :
$\zeta(0)=\zeta_{D}(0)+\zeta_{N}(0)=-{1\over 180}+{29\over 180}
={7\over 45}$.
For the spin-${1\over 2}$ field, we have found that a first-order
differential operator for the local boundary-value problem (1) exists which
is symmetric and has self-adjoint extensions [2,6]. A {\it direct}
calculation [2,6] gives :
$$
\zeta_{1\over 2}(0)={11\over 360} \; \; \; \; .
\eqno (2)
$$
For the spin-${3\over 2}$ field, it is not yet clear whether, and eventually
how, conditions (1) can be used to perform one-loop calculations.

An alternative set of local boundary conditions [7] is suggested by the
study of field transformation properties under local supersymmetry. These
involve the spatial components $\Bigr(\psi_{i}^{A}, \; {\widetilde \psi}_{i}
^{A'}\Bigr)$ of the spin-${3\over 2}$ potential, rather than the field
strength $\Bigr(\phi_{ABC}, \; {\widetilde \phi}_{A'B'C'}\Bigr)$.
In particular, in simple supergravity the spatial tetrad and
the projection $\Bigr(\pm {\widetilde \psi}_{i}^{A'}-\sqrt{2} \;
n_{A}^{\; \; A'}\psi_{i}^{A}\Bigr)$ transform into each other under half of
the local supersymmetry transformations at the boundary, so that they can
be specified as boundary data (up to gauge)
in computing the quantum amplitude [7]. Thus from this point of view the
most natural boundary conditions are Dirichlet for spin 2, and :
$\sqrt{2} \; n_{A}^{\; \; A'}\psi_{i}^{A} =\pm
{\widetilde \psi}_{i}^{A'}$ on $S^{3}$.
The PDF value for gravity is $\zeta(0)=-{278\over 45}$ in this case [3],
and for spin ${3\over 2}$ a {\it direct} $\zeta(0)$ calculation using
the above local boundary condition gives [2,6] :
$$
\zeta_{3\over 2}(0)=-{289\over 360} \; \; \; \; ,
\eqno (3)
$$
where $\zeta_{3\over 2}(0)$ is the PDF value, found using the gauge
condition $e_{AA'}^{\; \; \; \; \; j}\psi_{j}^{A}=0$. In the case of
$O(N)$ supergravity models, the remaining boundary conditions
consistent with supersymmetry transformation rules are Dirichlet
for scalar fields (and Neumann for pseudo-scalars), magnetic for
spin $1$, and the previous local boundary conditions (1) for massless
spin-${1\over 2}$ fields. One then finds [2] for the different $O(N)$
models with scale-invariant measures :
$$
\zeta_{T}^{(1)}(0)=-{43\over 8}, \;
\zeta_{T}^{(2)}(0)=-5 , \;
\zeta_{T}^{(3)}(0)=-{61\over 12}, \;
\zeta_{T}^{(4)}(0)=-{17\over 3},
\eqno (4)
$$
$$
\zeta_{T}^{(5)}(0)=-{41\over 6}, \;
\zeta_{T}^{(6)}(0)=-{55\over 6}, \;
\zeta_{T}^{(7)}(0)=\zeta_{T}^{(8)}(0)=-{83\over 6}.
\eqno (5)
$$
Thus $O(N)$ supergravity theories are not even one-loop finite in the
presence of boundaries, at least when the PDF values are used for gauge
fields. The
same conclusion has been reached in [5] using the Faddeev-Popov technique
and including the effect of antisymmetric tensor fields, which have not
been considered in deriving (4,5).

For fermionic fields, as we said at the beginning,
one has an additional choice of using non-local
(spectral) boundary conditions, rather than the previous local conditions.
This choice is not useful in studying
$O(N)$ supergravity models because it does not respect supersymmetry, but
it is nevertheless of some mathematical interest. For a
massless spin-${1\over 2}$ field, non-local boundary conditions require that
the part of the spin-${1\over 2}$ field with a particular sign of eigenvalues
of the intrinsic 3-dimensional Dirac operator should vanish
on $S^3$, and similarly for the spin-${3\over 2}$ potential.
This leads to [2] :
$$
\zeta_{1\over 2}(0)={11\over 360} \; \; \; \; , \; \; \; \;
\zeta_{3\over 2}(0)=-{289\over 360} \; \; \; \; ,
\eqno (6)
$$
where, again, $\zeta_{3\over 2}(0)$ is the PDF value. Remarkably, these are
equal to the local values (2,3).

A further topic for investigation is the relation between our PDF $\zeta(0)$
values for gauge fields and the results of BRST-invariant $\zeta(0)$
calculations including ghost fields.
\vskip 0.3cm
\leftline {\bf References}
\vskip 0.3cm
\item {[1]}
S. W. Hawking, Phys. Lett. B {\bf 126}, 175 (1983).
\item {[2]}
G. V. M. Esposito, Ph. D. Thesis, University of Cambridge, 1991.
\item {[3]}
K. Schleich, Phys. Rev. D {\bf 32}, 1889 (1985);
J. Louko, Phys. Rev. D {\bf 38}, 478 (1988).
\item {[4]}
I. G. Moss and S. Poletti, Nucl. Phys. B {\bf 341}, 155 (1990).
\item {[5]}
S. Poletti, Phys. Lett. B {\bf 249}, 249 (1990).
\item {[6]}
P. D. D'Eath and G. V. M. Esposito, Phys. Rev. D {\bf 43}, 3234 (1991).
\item {[7]}
H. C. Luckock and I. G. Moss, Class. Quantum Grav. {\bf 6}, 1993 (1989).
\bye